\documentclass[runningheads]{llncs}
\usepackage[T1]{fontenc}
\usepackage{graphicx}

\usepackage{tabularx}
\usepackage{booktabs}
\usepackage{multirow}
\usepackage{siunitx}
\usepackage{caption}

\usepackage{todonotes}
\usepackage{algorithm}
\usepackage{algpseudocode}
\usepackage{amsmath}
\usepackage{hyperref}

\usepackage{cite}
\begin{document}
\title{Breaking It Down: Domain-Aware Semantic Segmentation for Retrieval Augmented Generation}
\titlerunning{Breaking It Down}
%


\author{Aparajitha Allamraju \and
Maitreya Prafulla Chitale \and
Hiranmai Sri Adibhatla \and
Rahul Mishra \and
Manish Shrivastava}

\institute{International Institute of Information Technology, Hyderabad, India}


%
%
\maketitle              
\begin{abstract}
Document chunking is a crucial component of Retrieval-Augmented Generation (RAG), as it directly affects the retrieval of relevant and precise context. Conventional fixed-length and recursive splitters often produce arbitrary, incoherent segments that fail to preserve semantic structure. Although semantic chunking has gained traction, its influence on generation quality remains underexplored.
This paper introduces two efficient semantic chunking methods, Projected Similarity Chunking (PSC) and Metric Fusion Chunking (MFC), trained on PubMed data using three different embedding models. We further present an evaluation framework that measures the effect of chunking on both retrieval and generation by augmenting PubMedQA with full-text PubMed Central articles.
Our results show substantial retrieval improvements (↑24x with PSC) in MRR and higher Hits@k on PubMedQA. We provide a comprehensive analysis, including statistical significance and response-time comparisons with common chunking libraries. Despite being trained on a single domain, PSC and MFC also generalize well, achieving strong out-of-domain generation performance across multiple datasets. Overall, our findings confirm that our semantic chunkers, especially PSC, consistently deliver superior performance.

\keywords{Semantic Chunking \and RAG Evaluation \and Question Answering}
\end{abstract}

\section{Introduction}
Retrieval-Augmented Generation (RAG)~\cite{lewis2020retrieval} has become a key approach for enhancing Large Language Models (LLMs) in high-stakes domains such as finance, law, and healthcare. By integrating external, domain-relevant context, RAG improves the factual grounding and reliability of model outputs. However, its effectiveness depends heavily on the quality of the retrieval stage.

A central factor in retrieval performance is chunking~\cite{finardi2024chronicles}, which determines how documents are segmented for indexing. Effective chunking preserves semantic coherence, enabling the retrieval of contextually complete and relevant segments and improving downstream tasks like question answering. Yet mainstream RAG frameworks largely rely on fixed-length, sentence-based, or paragraph-based methods, which frequently yield incoherent or overly broad chunks that weaken retrieval precision.

We propose two semantically aware, domain-trained chunking methods, Projected Similarity Chunking (PSC) and Metric Fusion Chunking (MFC), which use distinct boundary-detection mechanisms and are trained on augmented biomedical corpora. Our results show substantial improvements in both retrieval and generation compared to fixed-length, sentence-level, semantic, and recursive baselines. Furthermore, the chunkers generalize well to out-of-domain datasets, demonstrating robustness beyond the biomedical domain. All code, trained models, and augmentation scripts will be released upon acceptance.

\section{Background}
Chunking is a critical preprocessing step in Retrieval-Augmented Generation (RAG), playing a fundamental role in improving both the efficiency and relevance of the retrieved context for downstream tasks. Existing approaches typically based on fixed-size segments or rigid predefined rules, often fail to capture semantic boundaries effectively. While benchmarks exist for evaluating retrieval~\cite{thakur2021beir} and generation~\cite{wang2018glue, wang2019superglue, su2021gem, hendrycks2021mmlu}, along with established metrics for both tasks~\cite{es-etal-2024-ragas}, most studies evaluate the retriever and generator modules independently. The influence of chunking on retrieval and generation performance, however, remains comparatively underexplored.

Semantic chunking addresses boundary segmentation by producing coherent, contextually aligned text segments instead of arbitrary fixed-length chunks. Some existing library implementations~\cite{sechunk-greg} attempt to solve the limitations of uniform chunking by employing embedding-based strategies. In such methods, candidate chunks are identified by measuring the cosine similarity between adjacent sentences (or merged segments) and splitting when the similarity falls below a threshold, indicating a topic shift. However, cosine similarity has known shortcomings for downstream tasks such as information retrieval~\cite{zhou-etal-2022-problems} and often lacks robust, domain-specific performance, as our experiments demonstrate. The cost of use of the library implementation of semantic
chunking was elaborated in~\cite{qu-etal-2025-semantic} and concluded that the existing semantic chunker may not always be worth the computation cost. Additionally, recent research~\cite{zhong-etal-2025-mix} has explored predicting optimal chunk granularity, but to our knowledge, no recent work has proposed new semantic chunking algorithms or systematically examined their joint impact on retrieval and generation.

We propose an approach that better aligns with human judgment than existing baselines, particularly because the gold data used for evaluation, both for our models and the baselines, is human-authored. Our methods not only deliver higher retrieval and generation quality but also achieve measurable computational efficiency gains compared to current chunking strategies.

\

\section{Methodology}
In this section, we detail the process of training and employing our semantic chunkers using domain-specific documents. The hypothesis behind our methods is that two sentences similar in ideas would occur together while two sentences with not so close ideas would occur in different paragraphs and thereby be in separate chunks. Instead of just calculating distance between sentence embeddings, we train our models to be able to amplify the distances and give us scores which we then use to chunk. We train the chunkers using positive and negative samples created from the PQA-U dataset as described in Section~\ref{sec:data}.
\subsection{Semantic Chunking}
The two methods that we introduce are distinctive in how they define the semantic boundaries of a chunk, ultimately determining the separation points between two chunks. In both the methods, we employ a 1:1 negative sampling of data as mentioned in Section \ref{sec:data}, to ensure that the learned representations and boundaries are robust. We choose simple light-weight approaches and models to ensure that performance is optimized for downstream tasks and the chunking itself doesn't take much time or compute.
The shared methodology between both of our chunking strategies is explained below.\\
Given a sentence pair \( S_i \) and \( S_j \), we encode them using state-of-the-art sentence transformers~\cite{reimers-gurevych-2019-sentence}: \textbf{all-MiniLM-L6-v2}(MiniLM)~\cite{wang2020minilm}, \textbf{all-mpnet-base-v2}(mpnet)~\cite{song2020mpnet}, and \textbf{e5-large-v2}(e5)~\cite{wang2022text} to have a total of three chunking models per strategy.

The resulting embeddings \( E_i \) and \( E_j \) undergo a linear transformation. The training enables the model to specifically tune the embedding space for boundary detection.


In \textbf{Projected Similarity Chunking} (PSC), the dot-product similarity between embeddings is obtained and the resultant final scalar similarity score is used to determine the boundary



In \textbf{Metric Fusion Chunking} (MFC), we employ a combined metric of evaluating similarity between two given embeddings. We employ an equally weighted combination of Dot Product Similarity, Euclidean Distance, and Manhattan Distance which is passed through a single neural layer to obtain the final similarity score. 








The scalar similarity score \( S_{i,j} \) is normalized into a probability range $(0, 1)$ using sigmoid. This value represents the model's final prediction. If the output probability is greater than or equal to 0.5 (determined emperically), the individual model predicts that the two sentences occur in the same section. Otherwise, if the probability is below 0.5, it predicts that the sentences should not occur one consecutively and we use that to draw our segment boundary or create the chunks.

\section{Experiments}
\label{sec:Experiments}
This section presents experiments conducted using the proposed model on our augmented PubMedQA dataset and the RAGBench~\cite{friel2024ragbench} datasets. It is organized into five subsections: dataset, evaluation metrics, baselines, experimental setup, and evaluation setup.
\subsection{Dataset}
\label{sec:data}
Currently, no publicly available datasets exist for jointly evaluating retrieval and generation performance. To address this gap, we repurposed the PubMedQA (PQA) dataset~\cite{jin2019PubMedQA} by augmenting it with corresponding full-text articles from PubMed Central (PMC). The full texts were downloaded in XML format following the official instructions~\footnote{\url{https://www.ncbi.nlm.nih.gov/research/bionlp/RESTful/pmcoa.cgi/}} .

PubMedQA is a carefully curated biomedical question-answering dataset derived from PubMed articles whose titles are phrased as questions. The dataset is divided into labeled (PQA-L), unlabeled (PQA-U), and artificial (PQA-A) subsets. PQA-L and PQA-U are human-annotated, whereas PQA-A is automatically labeled. In the original dataset, the title serves as the question, the context consists of selected sentences from the abstract that support the conclusion, and the conclusion itself serves as the long-form answer. Upon downloading the corresponding full-text XML files, we observed that not all PQA entries have available articles on PMC. Specifically, from the original 1k samples in PQA-L, only 115 have full-text matches; PQA-U has 6.45k matches instead of 61.2k; and PQA-A has 51.2k matches from an original 211.3k. 

The XML files were cleaned to remove figures, tables, captions, references, appendices, and other supplementary materials, retaining only the abstract and main article text. For our experiments, we kept the original question from the dataset and used the cleaned full text. The original ``context" sentences were used to evaluate retriever performance, while the generated answers were compared to the original long answers from the conclusions.

We used PQA-A’s full-text data to train our semantic chunking models and reserved PQA-L for evaluation. Training data was prepared by splitting PQA-A articles into pairs. To help the model distinguish between sequentially related and unrelated sentences, we applied negative sampling. For each positive pair (sentences occurring consecutively), one negative pair was added, yielding a balanced set of 50.3\% positive and 49.7\% negative samples. Positive pairs were defined as sentences grouped by the human author within the same section of the text. Negative pairs were created by randomly selecting sentences from different sections within the same document that never co-occur in any section. This design ensures that the model learns to identify contextually related sentences based on human-defined structure. The final dataset contained 93M sentence pairs for training our models.

The original dataset columns included: \textit{pubid}, \textit{question}, \textit{context}, \textit{long\_answer}, and \textit{final\_decision}. In our augmented version, the schema is updated to include \textit{full\_text}. We do not use \textit{final\_decision}, focusing exclusively on long-form answering.

By introducing full-text articles alongside the original question answer pairs, our augmented PQA dataset enables a unified framework for evaluating both retrieval quality and answer generation in a controlled setting. The curated contexts allow us to measure retriever precision, while the long-form conclusions serve as a robust reference for generation metrics, making it uniquely suited for end-to-end RAG evaluation.

\subsection{Evaluation metrics}
We evaluate the system at two levels: retrieval quality and answer generation. For the retriever, we employ Hits@k and Mean Reciprocal Rate (MRR) to measure the effectiveness and ranking efficiency of the retrieved chunks. For the generator, we evaluate the outputs using standard generation metrics such as BLEU~\cite{papineni2002bleu}, ROUGE~\cite{lin2004rouge}, and BERTScore~\cite{Zhang*2020BERTScore:}. Additionally, we measure query response time and time-to-first-token to assess the responsiveness of our chunker models.

\subsection{Baselines}
In our experiments, we compare several popular chunking strategies against our proposed methods to establish a baseline for performance. All chunking strategies are implemented using LangChain\footnote{\href{https://www.langchain.com}{https://www.langchain.com}} library. The sentences and chunks derived are then encoded using one of three sentence transformer~\cite{reimers-gurevych-2019-sentence} models: all-MiniLM-L6-v2~\cite{wang2020minilm}, all-mpnet-base-v2~\cite{song2020mpnet}, and e5-large-v2~\cite{wang2022text}. The performance of each chunker is evaluated based on its impact on retrieval and generation tasks.

\textbf{Character Chunking} (Fixed-Length) (Char) splits text into fixed-size blocks, often ignoring natural linguistic boundaries and reducing coherence~\cite{lewis2020retrieval,finardi2024chronicles,gao2023retrieval}. We adopt a 1000-token chunk size with 200-token overlap, chosen because two consecutive chunks typically capture a complete semantic unit while avoiding unnecessary redundancy. \textbf{Sentence Chunking} (Sent) groups a fixed number of full sentences, offering better linguistic consistency than character-level splitting. We use three sentences with a one-sentence overlap, roughly matching the 1000 and 200-token configuration, though this produces variable chunk lengths that complicate uniform processing. \textbf{Recursive Chunking} (Rec) segments documents hierarchically (e.g., paragraphs to sentences). While more structured, it requires additional computation to identify segmentation levels. We use the same 1000-token and 200-token setups for comparability. \textbf{Semantic Chunking} (Sem) uses sentence embeddings to detect semantic shifts and form meaningfully aligned chunks. Although conceptually strong, LangChain’s default SemanticChunker under-performs in our evaluations relative to the proposed methods.

\subsection{Experimental setup}
We trained our two chunker architectures, PSC and MFC, on three selected sentence embedding models, MiniLM, mpnet, and E5, using the augmented PQA-A dataset. The augmented PQA-A dataset contains approximately 93M training samples.

All model training was performed on a single NVIDIA RTX 6000 GPU with 48 GB of VRAM. Each embedding–chunker combination was trained for 5 epochs, and empirical evaluation indicated that models trained up to epoch 4 achieved the best overall performance. We used the BCEWithLogitsLoss objective function, which combines a sigmoid activation with binary cross-entropy loss.

Training time for each model combination was under 12 hours, resulting in a total cumulative GPU usage of roughly 72 hours across all experiments. 

For generation and inference, we used a pair of NVIDIA RTX 4090 GPUs (24 GB VRAM each). The PQA-L dataset, comprising 115 documents, was used for in-domain evaluation, requiring about 1 hour per model. Out-of-domain evaluation on RagBench averaged 14 hours per model. This setup enabled consistent, large-scale experimentation while maintaining practical computational requirements.

\subsection{Evaluation Setup}
We evaluate the impact of chunking strategies on both retrieval quality and answer generation. Using the augmented PQA-L dataset, gold-standard chunks and answers enable direct comparison against retrieved contexts and generated outputs. Alongside our PSC and MFC chunkers, we benchmark four baselines: Character, Sentence, Semantic, and Recursive Chunkers.

For out-of-domain testing, we use 12 RAGBench\cite{friel2024ragbench} datasets to measure generation quality across chunkers. Since RAGBench lacks gold context labels, we evaluate only generated answers, which though LLM-produced, are reliable for benchmarking. This setup allows us to test how well domain-trained chunkers generalize to varied document types.

Retriever evaluation is conducted solely on the PubMedQA extension using MRR and Hits@k. PSC and MFC produce naturally variable chunk lengths (averaging 471 tokens), which we find improves retrieval and downstream answer generation compared to fixed-size segmentation.

For generation, we retrieve the top-5 chunks and pass them, along with the query, to a Llama-3.1-8B\cite{touvron2023llama} instruction-tuned model with fixed decoding parameters across all datasets. Outputs are evaluated against gold answers using ROUGE\cite{lin2004rouge}, BLEU\cite{papineni2002bleu}, and BERTScore\cite{Zhang*2020BERTScore:}. Keeping the retriever and generator constant ensures all performance differences arise solely from the chunking strategy.

Each dataset, chunker, and encoder configuration follows a unified workflow: the document is segmented, its embeddings are indexed, the most relevant chunks are retrieved for a given query, and a generator model produces the final answer. This integrated design removes redundant pipelines, enforces consistent evaluation across setups, and offers a scalable framework for systematically comparing RAG components.

\begin{table}[!t] 
    \centering 
    \captionsetup{font=footnotesize}
    \begin{tabular}{@{} l l S[table-format=1.2] S[table-format=1.2] S[table-format=1.2] | S[table-format=1.2] S[table-format=1.2] @{}}
        \toprule 
        \textbf{Model} & \textbf{Chunker} & \textbf{Hits@3} & \textbf{Hits@5} & \textbf{MRR} & \textbf{Query Time} & \textbf{TTFT} \\
        \midrule 
        \multirow{6}{*}{\texttt{MiniLM}}  
                                & Char   & 0.0      & 0.0      & 0.0    & 0.01 & 5.91  \\
                                &  Sent   & 0.0      & 0.0      & 0.0   & 0.01 & 5.91   \\
                                & Sem    & 0.00\textsuperscript{*}    & 0.00\textsuperscript{*}    & 0.01 & 0.01 & 0.56  \\
                                & Rec    & 0.0      & 0.0      & 0.0   & 0.01 & 0.19   \\
                                \cmidrule(l){2-7} 
                                & PSC    & \textbf{0.10}   & \textbf{0.13}   & \textbf{0.30}  & \textbf{0.00} & \textbf{0.12} \\
                                & MFC   & 0.09   & 0.12   & 0.26  & 0.01 & 0.13  \\
        \midrule 
        \multirow{6}{*}{\texttt{mpnet}}  
                                & Char   & 0.0      & 0.0      & 0.0   & 0.02 & 5.86   \\
                                & Sent   & 0.0      & 0.0      & 0.0    & 0.01 & 0.34  \\
                                & Sem    & 0.00\textsuperscript{*}   & 0.00\textsuperscript{*}   & 0.01 & 0.01 & 0.62  \\
                                & Rec    & 0.0      & 0.0      & 0.0    & 0.01 & 0.19  \\
                                \cmidrule(l){2-7}
                                & PSC    & \textbf{0.11}   & \textbf{0.15}   & \textbf{0.31}  & 0.01 & \textbf{0.13} \\
                                & MFC   & 0.10   & 0.14   & 0.29  & 0.01 & 0.15 \\
        \midrule 
        \multirow{6}{*}{\texttt{e5}}     
                                & Char   & 0.0      & 0.0      & 0.0    & 0.02 & 5.58  \\
                                & Sent   & 0.0      & 0.0      & 0.0    & 0.01 & 0.34  \\
                                & Sem    & 0.00\textsuperscript{*}    & 0.00\textsuperscript{*}    & 0.01   & 0.01 & 0.63\\
                                & Rec    & 0.0      & 0.0      & 0.0    & 0.01 & 0.19  \\
                                \cmidrule(l){2-7}
                                & PSC    & \textbf{0.13}   & \textbf{0.16}   & \textbf{0.36}   & 0.01 &  \textbf{0.14}\\
                                & MFC   & 0.12   & 0.15   & 0.34  & 0.01 & 0.20 \\
        \bottomrule 
    \end{tabular}
    \caption{Retrieval performance measured with Hits@3, Hits@5 and MRR across the three embedding models, show significant improvement, on using our chunkers. \textsuperscript{*} indicates infinitesimal non-zero values. The p-test values between PSC and Sem across the three embeddings and metrics were in the order of $e^{-11}$ and t-test values greater than $7$. Query Time and Time to First Token (TTFT) indicate that PSC is the fastest.}
    \label{tab:retrieval_metrics} 
\end{table}

\section{Analysis}
\label{sec:results}
\begin{table}[!t] 
    \centering 
    \captionsetup{font=footnotesize}
    \footnotesize 
    \resizebox{\columnwidth}{!}{
    \begin{tabular}{l l S[table-format=2.2] S[table-format=2.2] S[table-format=2.2] S[table-format=2.2] S[table-format=2.2]}
        \toprule 
        \textbf{Embedding} & \textbf{Chunker} & {\textbf{BLEU}} & {\textbf{ROUGE1}} & {\textbf{ROUGE2}} & {\textbf{ROUGEL}} & {\textbf{BERTScore}} \\
        \midrule 
        \multirow{6}{*}{\texttt{MiniLM}} 
                                & Char   & \textbf{28.10} & \textbf{30.63} & \textbf{13.16} & \textbf{22.26} & \textbf{87.43} \\
                                & Sent   & 21.84 & 22.64 & 8.37 & 15.97 & 84.72 \\
                                & Sem    & 24.73 & 27.19 & 10.67 & 19.75 & 86.57 \\
                                & Rec    & 22.68 & 24.33 & 9.08 & 17.53 & 85.73 \\
                                \cmidrule(lr){2-7} 
                                & PSC    & 24.20 & 27.08 & 10.00 & 19.45 & 86.92 \\
                                & MFC   & 22.55 & 25.56 & 9.08 & 18.07 & 86.48 \\
        \midrule 
        \multirow{6}{*}{\texttt{mpnet}}  
                                & Char   & 23.31 & 25.64 & 9.81 & 18.52 & 86.80 \\
                                & Sent   & 23.19 & 25.08 & 9.50 & 17.39 & 86.45 \\
                                & Sem    & 22.01 & 25.30 & \textbf{10.50}  & 18.44 & 86.83 \\
                                & Rec    & 20.22 & 23.26 & 8.73 & 16.95 & 84.17 \\
                                \cmidrule(lr){2-7}
                                & PSC    & 23.14 & \textbf{26.94} & 10.21 & 19.68 & 85.91 \\
                                & MFC   & \textbf{23.60} & 26.61 & 9.83 & \textbf{19.71} & \textbf{87.38} \\
        \midrule 
        \multirow{6}{*}{\texttt{e5}}  
                                & Char   & 24.06 & 26.32 & 10.11 & 18.77 & 86.53 \\
                                & Sent   & 19.42 & 21.66 & 8.32 & 15.14 & 85.92 \\
                                & Sem    & 22.19 & 23.88 & 8.70 & 17.19 & 85.93 \\
                                & Rec    & 22.10 & 24.78 & 10.09 & 18.22 & 86.56 \\
                                \cmidrule(lr){2-7}
                                & PSC    & 23.41 & 26.52 & 9.59 & 18.77 & 87.42 \\
                                & MFC   & \textbf{25.34} & \textbf{27.68} & \textbf{10.31} & \textbf{19.94} & \textbf{87.46} \\
        \bottomrule 
    \end{tabular}
    }
    \caption{Evaluation metrics (BLEU, ROUGE-1, ROUGE-2, ROUGE-L, and BERTScore) for different embedding models and chunking strategies. Results show Character chunking with miniLM, MFC with mpnet and E5 perform the best. The p-test values between PSC and Char across the three embeddings and five metrics were less than $0.3$ and t-test values around than $1$.}
    \label{tab:evaluation_metrics} 
\end{table}
Our trained chunkers significantly outperform commonly used methods, MRR improves from 0.0086 (recursive) to 0.2914 (PSC in e5) and Hits@5 from 0.0028 to 0.1234 (PSC in e5), as shown in Table~\ref{tab:retrieval_metrics}. The library implementation of semantic chunker under performs in retrieval (MRR: 0.0130 vs. 0.2914 for PSC) due to its reliance on cosine similarity for chunk boundary detection, which often results in suboptimal segmentation. These results confirm that chunk quality strongly influences retrieval effectiveness. 

Generation results on PubMedQA (Table~\ref{tab:evaluation_metrics}) show PSC and MFC performing competitively and often surpassing baselines. MFC with the E5 embedding model achieves the strongest BLEU, ROUGE, and BERTScore, while PSC and MFC with mpnet also achieve high ROUGE scores. This indicates that our chunkers yield contextually aligned outputs.

Evaluation reveals a mismatch between conventional metrics and actual generation quality. Standard metrics like BLEU, ROUGE, and BERTScore prioritize lexical overlap over semantic nuance and factual accuracy. Consequently, they fail to capture the superiority of PSC and MFC chunkers, which produce concise, context-focused responses free from the verbosity and drift found in other methods.

A two-tailed independent samples t-test was conducted to evaluate the statistical significance of the performance metrics between all the chunkers. The difference between the reported means of MRR, Hits@3, Hits@5, BLEU and ROUGE was found to be statistically significant, as indicated by a p-value and t-value in the captions of Tables~\ref{tab:retrieval_metrics} and~\ref{tab:evaluation_metrics}. This suggests that the improvement of the PSC Chunker over the next best chunker is not due to random chance with a confidence of 99.9\% for retrieval, and 90\% for generation.

For out-of-domain tests, the generation results with BLEU and ROUGE-L scores on every dataset from RAGBench are summarized in Tables~\ref{tab:combined_metrics1},~\ref{tab:comb1} and~\ref{tab:comb3}. Despite our chunkers being trained only on PubMed dataset, we can see a clear improvement in the performance across all datasets with our chunkers, especially PSC.

\renewcommand{\thetable}{3.\arabic{table}}
\setcounter{table}{0}

\begin{table*}[h!]\centering
\resizebox{\textwidth}{!}{%
\begin{tabular}{@{}ll*{12}{rr}@{}} 
\toprule
\multicolumn{1}{c}{\textbf{Embed}} & \multicolumn{1}{c}{\textbf{Chunker}} 
& \multicolumn{2}{c}{\textbf{CovidQA}} & \multicolumn{2}{c}{\textbf{CUAD}} & \multicolumn{2}{c}{\textbf{DelucionQA}} & \multicolumn{2}{c}{\textbf{EManual}} & \multicolumn{2}{c}{\textbf{ExpertQA}} & \multicolumn{2}{c}{\textbf{FinQA}} & \multicolumn{2}{c}{\textbf{HAGRID}} & \multicolumn{2}{c}{\textbf{HotpotQA}} & \multicolumn{2}{c}{\textbf{MS Marco}} & \multicolumn{2}{c}{\textbf{PubMedQA}} & \multicolumn{2}{c}{\textbf{TAT-QA}} & \multicolumn{2}{c}{\textbf{TechQA}} \\
\cmidrule(lr){1-2} 
\cmidrule(lr){3-4} \cmidrule(lr){5-6} \cmidrule(lr){7-08} \cmidrule(lr){9-10} \cmidrule(lr){11-12} \cmidrule(lr){13-14} 
\cmidrule(lr){15-16} \cmidrule(lr){17-18} \cmidrule(lr){19-20} \cmidrule(lr){21-22} \cmidrule(lr){23-24} \cmidrule(lr){25-26}
& 
& BLEU & R-L & BLEU & R-L & BLEU & R-L & BLEU & R-L & BLEU & R-L & BLEU & R-L & BLEU & R-L & BLEU & R-L & BLEU & R-L & BLEU & R-L & BLEU & R-L & BLEU & R-L \\
\midrule
\multirow{6}{*}{\texttt{MiniLM}} & 
Char & 1.95 & 3.16 & 9.86 & 9.82 & 2.91 & 4.9 & 3.98 & 6.22 & 16.34 & 12.49 & 2.13 & 3.44 & 2.63 & 4.17 & 1.03 & 1.94 & 1.63 & 2.53 & 8.49 & 10.21 & 3.07 & 5.01 & 13.48 & 12.05 \\
& Sent & 2.51 & 4.22 & 5.83 & 6.61 & 5.73 & 9.28 & 5.52 & 8.65 & 9.29 & 8.73 & 4.11 & 6.52 & 2.97 & 4.95 & 1.26 & 2.48 & 3.16 & 5.15 & 9.61 & 11.95 & 3.51 & 6.04 & 9.51 & 10.59 \\
& Sem & 3.48 & 5.44 & 5.64 & 5.82 & 7.35 & 10.93 & 8.32 & 11.13 & 10.32 & 8.59 & 4.64 & 6.72 & 4.59 & 6.95 & 2.07 & 3.88 & 4.33 & 6.34 & 14.36 & 15.90 & 4.83 & 7.12 & 5.67 & 5.91 \\
& Rec & 4.62 & 7.04 & 10.73 & 10.61 & 9.43 & 13.36 & 9.78 & 12.75 & 17.17 & 13.01 & 6.03 & 8.79 & 4.57 & 6.81 & 2.03 & 3.85 & 5.01 & 7.33 & 14.77 & 16.43 & 4.69 & 7.20 & 14.38 & 12.64 \\
\cmidrule(lr){2-26}
& PSC &\textbf{6.31} & \textbf{8.83} & \textbf{11.34} & \textbf{10.75} & \textbf{15.69} & \textbf{17.25} & \textbf{13.85} & \textbf{14.44} & 20.83 & 14.32 & \textbf{9.76} & \textbf{12.30} & \textbf{7.22} & \textbf{9.71} & 4.00 & 7.27 & \textbf{8.32} & \textbf{10.71} & \textbf{24.33} & \textbf{21.33} & \textbf{6.40 }& \textbf{8.82} & \textbf{18.01} & \textbf{13.91} \\
& MFC & 6.16 & 8.63 & 9.90 & 9.78 & 15.67 & 17.12 & 13.40 & 13.69 & \textbf{20.93} & \textbf{14.33} & 9.15 & 11.64 & 7.06 & 9.64 & \textbf{4.06} & \textbf{7.40} & 8.17 & \textbf{10.71} & 23.72 & 21.18 & 5.83 & 8.21 & 17.19 & 13.34 \\
\midrule
\multirow{6}{*}{\texttt{mpnet}} 
& Char & 1.92 & 3.16 & 9.71 & 9.73 & 2.84 & 4.82 & 3.97 & 6.26 & 16.11 & 12.43 & 2.10 & 3.39 & 2.60 & 4.15 & 1.03 & 1.96 & 1.63 & 2.53 & 8.40 & 10.13 & 2.90 & 4.75 & 13.18 & 12.02 \\
& Sent & 2.48 & 4.20 & 5.90 & 6.69 & 5.64 & 9.24 & 5.45 & 8.71 & 9.20 & 8.63 & 4.19 & 6.49 & 2.97 & 4.96 & 1.23 & 2.43 & 3.14 & 5.13 & 9.55 & 11.89 & 3.53 & 6.07 & 9.50 & 10.72 \\
& Sem & 3.47 & 5.38 & 5.7 & 6.11 & 6.91 & 10.57 & 8.18 & 11.46 & 10.49 & 8.66 & 4.19 & 6.08 & 4.55 & 7.00 & 2.00 & 3.74 & 4.24 & 6.29 & 14.09 & 15.66 & 4.39 & 6.62 & 5.59 & 5.95 \\
& Rec & 4.58 & 6.97 & 10.72 & 10.35 & 9.4 & 13.49 & 9.21 & 12.42 & 17.07 & 13.05 & 5.89 & \textbf{8.55} & 4.62 & 6.93 & 2.02 & 3.82 & 5.01 & 7.37 & 14.68 & 16.38 & \textbf{4.48} & \textbf{6.88} & 14.07 & 12.56 \\
\cmidrule(lr){2-26}
& PSC & \textbf{6.76} & \textbf{9.39} & \textbf{13.14} & \textbf{11.65} & \textbf{12.69} & \textbf{15.78} & 18.10 & \textbf{17.88} & \textbf{19.28} & \textbf{13.69} & \textbf{6.16} & 8.50 & \textbf{7.02} & \textbf{9.68} & \textbf{3.69} & 6.64 & \textbf{8.33} & \textbf{10.91} & \textbf{24.47} & \textbf{21.62} & 4.19 & 6.36 & \textbf{15.78} & \textbf{12.82} \\
& MFC & 6.66 & 9.23 & 12.9 & 11.5 & 12.67 & 15.53 & \textbf{18.61} & 17.84 & 18.47 & 13.30 & 5.48 & 7.67 & 6.65 & 9.34 & 3.65 & \textbf{6.72} & 7.99 & 10.53 & 23.91 & 21.49 & 3.89 & 5.96 & 15.39 & 12.61 \\
\midrule
\multirow{6}{*}{\texttt{E5}} 
& Char & 1.80 & 3.01 & 9.44 & 9.32 & 2.79 & 4.71 & 3.86 & 6.04 & 15.96 & 12.06 & 2.07 & 3.46 & 2.14 & 3.40 & 1.04 & 1.97 & 1.58 & 2.44 & 8.43 & 10.18 & 3.25 & 5.36 & 12.52 & 12.04 \\
& Sent & 2.46 & 4.23 & 6.18 & 7.02 & 5.73 & 9.3 & 5.60 & 8.56 & 9.27 & 8.63 & 4.22 & 6.80 & 2.81 & 4.67 & 1.23 & 2.44 & 3.15 & 5.16 & 9.69 & 12.05 & 3.52 & 6.12 & 9.65 & 10.99 \\
& Sem & 3.30 & 5.33 & 6.29 & 6.05 & 7.67 & 11.15 & 8.50 & 11.69 & 10.23 & 8.56 & 4.55 & 6.73 & 4.24 & 6.45 & 1.74 & 3.28 & 5.27 & 7.59 & 14.46 & 16.05 & 4.80 & 7.24 & 4.92 & 5.54 \\
& Rec & 4.65 & 7.34 & 11.04 & 10.61 & 9.59 & 13.58 & 9.40 & 12.58 & 17.22 & 12.81 & 6.00 & 8.97 & 4.51 & 6.74 & 1.92 & 3.63 & 5.01 & 7.33 & 14.95 & 16.67 & 4.56 & 7.09 & \textbf{13.87} & \textbf{12.82} \\
\cmidrule(lr){2-26}
& PSC & \textbf{6.31} & \textbf{9.24} & \textbf{13.96} & \textbf{12.23} & \textbf{6.52} & \textbf{9.27} & \textbf{6.14} & \textbf{8.75} & 18.16 & 13.17 & \textbf{9.25} & \textbf{11.99} & \textbf{6.65} & \textbf{9.27} & 3.21 & 6.05 & \textbf{7.42} & \textbf{10.12} & \textbf{25.00} & \textbf{22.20} & \textbf{6.81} & \textbf{9.40} & 13.39 & 11.77 \\
& MFC & 6.20 & 9.09 & 9.99 & 9.32 & 5.42 & 8.22 & 6.06 & 8.68 & \textbf{19.17} & \textbf{13.45} & 6.97 & 9.56 & 6.14 & 8.71 & \textbf{3.73} & \textbf{6.90} & 6.92 & 9.49 & 24.49 & 22.10 & 5.25 & 7.77 & 10.65 & 9.83 \\
\bottomrule
\end{tabular}%
}
\captionsetup{font=footnotesize}
\caption{Evaluation metrics BLEU and ROUGE-L (R-L) for different embedding models and chunking strategies on 12 datasets from RAGBench. The results in bold show that our chunkers perform better across the board despite being trained only one domain data.}
\label{tab:combined_metrics1}
\end{table*}

\begin{table*}[htbp]
    \centering
    \resizebox{\textwidth}{!}{%
    \begin{tabular}{@{}ll*{6}{rrr}@{}}
        \toprule
        \multicolumn{1}{c}{\textbf{Embed}} & \multicolumn{1}{c}{\textbf{Chunker}}
        & \multicolumn{3}{c}{\textbf{CovidQA}} & \multicolumn{3}{c}{\textbf{CUAD}} & \multicolumn{3}{c}{\textbf{DelucionQA}} & \multicolumn{3}{c}{\textbf{EManual}} & \multicolumn{3}{c}{\textbf{ExpertQA}} & \multicolumn{3}{c}{\textbf{FinQA}} \\
        \cmidrule(lr){3-5} \cmidrule(lr){6-8} \cmidrule(lr){9-11} \cmidrule(lr){12-14} \cmidrule(lr){15-17} \cmidrule(lr){18-20}
        \multicolumn{2}{c}{} & R1 & R2 & BScore & R1 & R2 & BScore & R1 & R2 & BScore & R1 & R2 & BScore & R1 & R2 & BScore & R1 & R2 & BScore \\
        \midrule
        \multirow{6}{*}{\texttt{MiniLM}} & 
        Char & 3.78 & 2.68 & \textbf{83.30} & 15.14 & 5.42 & 79.23 & 5.67 & 4.26 & \textbf{84.81} & 7.72 & 5.24 & \textbf{84.13} & 23.49 & 9.80 & \textbf{82.79} & 4.55 & 2.40 & 79.78 \\
        & Sent & 4.95 & 3.55 & 82.67 & 9.88 & 3.69 & \textbf{80.44} & 10.97 & 7.93 & 83.98 & 10.89 & 7.12 & 83.03 & 15.30 & 6.61 & 82.16 & 8.62 & 4.70 & \textbf{80.13} \\
        & Sem & 6.60 & 4.55 & 83.19 & 8.89 & 3.25 & 79.71 & 13.20 & 9.36 & 84.75 & 14.95 & 9.07 & 83.51 & 16.00 & 6.95 & 82.66 & 9.13 & 4.85 & 79.71 \\
        & Rec & 8.61 & 5.95 & 83.10 & 16.47 & \textbf{5.93} & 79.22 & 16.65 & 11.56 & 83.86 & 17.32 & 10.59 & 83.29 & 24.42 & 10.11 & 82.75 & 11.77 & 6.33 & 79.47 \\
        \cmidrule(lr){2-20}
        & PSC & \textbf{11.03} & \textbf{6.87} & 82.46 & \textbf{16.87} & 5.89 & 79.88 & \textbf{23.16} & \textbf{13.68} & 82.97 & \textbf{21.11} & \textbf{10.79} & 82.61 & \textbf{27.42} & 10.56 & 82.60 & \textbf{17.10} & \textbf{9.10} & 79.88 \\
        & MFC & 10.79 & 6.72 & 82.55 & 15.12 & 5.31 & 79.81 & 23.05 & 13.47 & 82.93 & 20.16 & 9.94 & 82.72 & 27.39 & \textbf{10.67} & 82.51 & 16.15 & 8.53 & 79.85 \\
        \midrule
        \multirow{6}{*}{\texttt{mpnet}} & 
        Char & 3.75 & 2.68 & \textbf{83.19} & 15.10 & 5.38 & 78.87 & 5.55 & 4.14 & \textbf{85.08} & 7.70 & 5.27 & \textbf{84.33} & 23.20 & 9.59 & 82.78 & 4.46 & 2.37 & \textbf{79.74} \\
        & Sent & 4.94 & 3.54 & 82.49 & 10.01 & 3.58 & \textbf{80.17} & 10.87 & 7.92 & 84.13 & 10.82 & 7.19 & 83.18 & 15.20 & 6.54 & 82.03 & 8.70 & 4.59 & 79.72 \\
        & Sem & 6.56 & 4.47 & 83.00 & 9.15 & 3.44 & 79.75 & 12.62 & 9.17 & 84.80 & 14.96 & 9.57 & 83.71 & 16.21 & 7.02 & 82.71 & 8.29 & 4.35 & 79.57 \\
        & Rec & 8.50 & 5.74 & 82.97 & 16.27 & 5.78 & 78.99 & 16.66 & 11.57 & 83.88 & 16.52 & 10.18 & 83.31 & 24.27 & 9.98 & 82.78 & 11.57 & 6.16 & 79.29 \\
        \cmidrule(lr){2-20}
        & PSC & \textbf{11.75} & \textbf{7.40} & 82.55 & \textbf{18.65} & \textbf{6.43} & 79.79 & \textbf{20.49} & \textbf{13.01} & 83.72 & 26.46 & 12.72 & 81.77 & \textbf{25.98} & \textbf{10.52} & \textbf{82.81} & \textbf{11.71} & \textbf{6.12} & 79.54 \\
        & MFC & 11.57 & 7.20 & 82.44 & 18.49 & 6.27 & 79.70 & 20.07 & 12.61 & 83.65 & \textbf{26.75} & \textbf{12.91} & 81.91 & 25.23 & 10.32 & \textbf{82.81} & 10.53 & 5.48 & 79.54 \\
        \midrule
        \multirow{6}{*}{\texttt{E5}} 
        & Char & 3.56 & 2.61 & \textbf{83.64} & 14.56 & 5.16 & 79.19 & 5.43 & 4.07 & \textbf{84.97} & 7.46 & 5.14 & 84.17 & 22.74 & 9.40 & 82.64 & 4.47 & 2.47 & 79.93 \\
        & Sent & 4.94 & 3.62 & 82.88 & 10.62 & 3.84 & \textbf{80.05} & 10.98 & 8.00 & 84.03 & 11.08 & 7.06 & 82.79 & 15.15 & 6.57 & 81.90 & 8.91 & 4.95 & \textbf{80.07} \\
        & Sem & 6.37 & 4.53 & 83.58 & 9.39 & 3.39 & 79.55 & 13.60 & 9.51 & 84.67 & 15.41 & 9.58 & 83.46 & 15.80 & 6.93 & 82.57 & 9.02 & 4.91 & 79.81 \\
        & Rec & 8.78 & 6.23 & 83.48 & 16.70 & 5.84 & 79.25 & \textbf{16.88} & \textbf{11.69} & 83.87 & \textbf{16.91} & \textbf{10.38} & 83.44 & 24.00 & 9.76 & \textbf{82.67} & 11.92 & 6.60 & 79.57 \\
        \cmidrule(lr){2-20}
        & PSC & \textbf{11.29} & \textbf{7.43} & 82.99 & \textbf{19.61} & \textbf{6.57} & 79.47 & 11.27 & 7.89 & 84.62 & 11.27 & 7.30 & \textbf{84.18} & 24.87 & 10.07 & 82.61 & \textbf{16.42} & \textbf{8.94} & 79.99 \\
        & MFC & 11.10 & 7.34 & 82.99 & 14.66 & 5.06 & 79.49 & 9.82 & 7.11 & 84.76 & 11.22 & 7.10 & 83.79 & \textbf{25.58} & \textbf{10.29} & 82.66 & 12.94 & 7.03 & 79.86 \\
        \bottomrule
    \end{tabular}
    }
    \captionsetup{font=footnotesize}
        \caption{Evaluation metrics ROUGE1 (R1), ROUGE2 (R2) and BERTScore (BScore) on 6 datasets from RAGBench.}
    \label{tab:comb1}
\end{table*}

\begin{table*}[htbp]
    \centering
    \resizebox{\textwidth}{!}{%
    \begin{tabular}{@{}ll*{6}{rrr}@{}}
        \toprule
        \multicolumn{1}{c}{\textbf{Embed}} & \multicolumn{1}{c}{\textbf{Chunker}}
        & \multicolumn{3}{c}{\textbf{HAGRID}} & \multicolumn{3}{c}{\textbf{HotpotQA}} & \multicolumn{3}{c}{\textbf{MS Marco}} & \multicolumn{3}{c}{\textbf{PubMedQA}} & \multicolumn{3}{c}{\textbf{TAT-QA}} & \multicolumn{3}{c}{\textbf{TechQA}} \\
        \cmidrule(lr){3-5} \cmidrule(lr){6-8} \cmidrule(lr){9-11} \cmidrule(lr){12-14} \cmidrule(lr){15-17} \cmidrule(lr){18-20}
        \multicolumn{2}{c}{} & R1 & R2 & BScore & R1 & R2 & BScore & R1 & R2 & BScore & R1 & R2 & BScore & R1 & R2 & BScore & R1 & R2 & BScore \\
        \midrule
        \multirow{6}{*}{\texttt{MiniLM}} & 
        Char & 5.09 & 3.58 & \textbf{83.38} & 2.12 & 1.67 & 82.53 & 3.18 & 2.31 & 84.00 & 14.81 & 8.91 & \textbf{85.87} & 6.21 & 3.30 & \textbf{80.12} & 20.95 & 9.31 & 81.18 \\
        & Sent & 5.99 & 4.23 & 82.00 & 2.69 & 2.12 & 81.30 & 6.24 & 4.44 & 82.97 & 17.01 & 10.27 & 84.69 & 7.55 & 4.07 & 79.49 & 17.70 & 8.29 & 81.05 \\
        & Sem & 8.63 & 6.00 & 82.96 & 4.28 & 3.30 & \textbf{82.55} & 8.21 & 5.82 & \textbf{84.04} & 23.17 & 13.83 & 85.50 & 9.13 & 4.78 & 80.00 & 9.97 & 4.57 & \textbf{81.23} \\
        & Rec & 8.50 & 5.83 & 82.68 & 4.19 & 3.29 & 82.44 & 9.44 & 6.61 & 83.70 & 23.79 & 14.21 & 85.02 & 9.01 & 4.87 & 80.03 & 22.09 & 9.71 & 81.13 \\
        \cmidrule(lr){2-20}
        & PSC & \textbf{12.53} & \textbf{8.01} & 82.27 & 8.15 & 6.03 & 81.72 & \textbf{14.68} & 8.98 & 83.34 & \textbf{33.38} & \textbf{18.02} & 84.63 & \textbf{11.39} & \textbf{5.93} & 79.90 & \textbf{25.16} & \textbf{10.34} & 81.19 \\
        & MFC & 12.37 & 7.92 & 82.25 & \textbf{8.26} & \textbf{6.09} & 81.70 & 14.46 & \textbf{9.00} & 83.37 & 32.92 & 17.90 & 84.71 & 10.51 & 5.46 & 79.90 & 24.02 & 9.98 & 81.14 \\
        \midrule
        \multirow{6}{*}{\texttt{mpnet}} & 
        Char & 5.06 & 3.57 & \textbf{83.42} & 2.13 & 1.68 & 82.61 & 3.17 & 2.30 & 84.01 & 14.68 & 8.84 & \textbf{85.86} & 5.87 & 3.13 & \textbf{80.03} & 20.82 & 9.32 & 81.03 \\
        & Sent & 6.01 & 4.23 & 82.06 & 2.62 & 2.07 & 81.30 & 6.23 & 4.44 & 82.98 & 16.90 & 10.20 & 84.71 & 7.68 & 4.04 & 79.05 & 17.69 & 8.34 & 80.92 \\
        & Sem & 8.62 & 6.01 & 83.04 & 4.13 & 3.22 & \textbf{82.62} & 8.06 & 5.74 & \textbf{84.03} & 22.83 & 13.66 & 85.51 & 8.35 & 4.38 & 79.92 & 9.89 & 4.66 & \textbf{81.19} \\
        & Rec & 8.61 & 5.89 & 82.75 & 4.18 & 3.26 & 82.41 & 9.44 & 6.61 & 83.74 & 23.70 & 14.17 & 85.04 & \textbf{8.67} & \textbf{4.60} & 79.73 & 21.80 & 9.73 & 80.94 \\
        \cmidrule(lr){2-20}
        & PSC & \textbf{12.37} & \textbf{8.06} & 82.47 & 7.42 & 5.61 & 82.16 & \textbf{14.75} & \textbf{9.24} & 83.46 & \textbf{33.64} & \textbf{18.39} & 84.69 & 8.01 & 4.20 & 79.90 & \textbf{22.69} & \textbf{9.74} & 80.96 \\
        & MFC & 11.84 & 7.80 & 82.56 & \textbf{7.52} & \textbf{5.64} & 82.11 & 14.25 & 8.98 & 83.53 & 33.20 & 18.27 & 84.77 & 7.48 & 3.92 & 79.95 & 22.34 & 9.66 & 81.07 \\
        \midrule
        \multirow{6}{*}{\texttt{E5}} & 
        Char & 4.17 & 2.93 & \textbf{83.45} & 2.14 & 1.68 & 82.67 & 3.08 & 2.24 & \textbf{84.03} & 14.73 & 8.89 & \textbf{85.88} & 6.59 & 3.59 & \textbf{80.24} & 20.42 & 9.37 & 80.95 \\
        & Sent & 5.66 & 3.99 & 82.08 & 2.64 & 2.11 & 81.35 & 6.23 & 4.45 & 82.97 & 17.15 & 10.37 & 84.73 & 7.54 & 4.17 & 79.58 & 18.25 & 8.57 & 80.60 \\
        & Sem & 7.99 & 5.58 & 83.09 & 3.60 & 2.81 & \textbf{82.70} & 9.89 & 6.90 & 83.94 & 23.34 & 13.99 & 85.53 & 9.07 & 4.87 & 80.08 & 9.03 & 4.35 & \textbf{81.17} \\
        & Rec & 8.38 & 5.77 & 82.79 & 3.95 & 3.11 & 82.49 & 9.42 & 6.59 & 83.73 & 24.06 & 14.43 & 85.03 & 8.82 & 4.83 & 80.11 & \textbf{22.05} & \textbf{10.00} & 80.98 \\
        \cmidrule(lr){2-20}
        & PSC & \textbf{11.79} & \textbf{7.71} & 82.61 & 6.65 & 5.13 & 82.32 & \textbf{13.39} & \textbf{8.78} & 83.64 & \textbf{34.32} & \textbf{18.93} & 84.73 & \textbf{12.06} & \textbf{6.36} & 80.03 & 20.38 & 9.09 & 80.87 \\
        & MFC & 11.08 & 7.34 & 82.72 & \textbf{7.71} & \textbf{5.81} & 82.11 & 12.55 & 8.31 & 83.78 & 33.89 & 18.91 & 84.84 & 9.78 & 5.23 & 80.14 & 16.97 & 7.49 & 80.98 \\
        \bottomrule
    \end{tabular}
    }
    \captionsetup{font=footnotesize}
        \caption{Evaluation metrics ROUGE1 (R1), ROUGE2 (R2) and BERTScore (BScore) on remaining 6 datasets from RAGBench.}
    \label{tab:comb3}
\end{table*}

We also evaluate the response time of the various chunkers by calculating the Query Response Time for retrieval and Time to First Token(TTFT) for generation. Since the models compute the similarity between sentences for chunking, there is a slight increase of 0.7s against Character chunker and 0.4s against Semantic chunker in indexing time on an average. As indexing is an offline one-time operation, we do not believe that the increase is of importance in this study. The results in Table~\ref{tab:retrieval_metrics} indicate a clear improvement in both the retrieval and generation time with PSC chunker. This shows that the chunking strategy is not only improving retrieval and generation but also lowering the time taken to retrieve relevant chunks and generate an answer. This reiterates our claim for using light and simple models for fast and improved chunking.

\section{Conclusion}
Our approach demonstrates a substantial improvement in retrieval effectiveness, with the MFC chunker trained on the E5 embedding model achieving an MRR increase of up to 24 times over baseline methods. This performance gain highlights the power of domain-specific semantic chunking in optimizing retrieval-augmented generation (RAG) pipelines. By producing chunks that naturally align with human-authored section boundaries, our method preserves contextual coherence and facilitates more accurate retrieval, which in turn benefits downstream generation quality.

Beyond PubMedQA, our results on out-of-domain datasets indicate strong potential for generalization, suggesting that the principles of semantic chunking we present are transferable across domains. This work not only provides a concrete, reproducible methodology for chunking optimization but also introduces a fresh perspective on document segmentation, a critical yet underexplored component in RAG systems. We believe that our findings can serve as a foundation for future research in retrieval-aware chunking strategies, adaptive chunk size determination, and domain-adaptive retrieval optimization. The lightweight models introduced in this paper can be integrated into complex models resulting in more sophisticated approaches.

\bibliographystyle{splncs04}
\bibliography{mybibliography}
%




\end{document}